# Supergravitational Production of Flavor and Color

J. Towe

*The Antelope Valley College, 3041 West Avenue K, Lancaster, CA 93536-5426, USA*
jtowe@avc.edu

A conventional graviton vertex operator is generated from the gauging of global supersymmetry. If the outgoing gravitino is replaced by a fermion-boson pair of like helicity, and if a new principle of equivalence associates supergravity with the introduction of color, then every quark flavor can be described as a gravitationally excited lepton. In this context, a SUSY SU(5) symmetry can be constructed exclusively of leptons and their scaler superpartners (or of quarks and their superpartners). This SUSY GUT flavor symmetry, which is preserved by the proposed supergravity interactions, accounts for three fermionic generations and predicts a new quark—a left-handed (non-strange) version of the strange quark.

## I. The Graviton Vertex Operator

The action for local supersymmetry is given by

$$S = -(1/2\kappa^2) \int d^4x \, |\det e| \, R - (1/2) \int d^4x \, \varepsilon^{\mu\nu\rho\sigma} \, \overline{\psi}_\mu \gamma_5 \gamma_\nu \overline{D}_\rho \Psi_\sigma, \qquad (1)$$

which is the sum of the Einstein action (R is the curvature scaler) and the action of the Rarita-Schwinger field, covariantized in terms of the covariant derivative

$$D_\mu = \partial_\mu - i \, \overline{\omega}_{\mu mn}(\sigma^{mn}/4). \qquad (2)$$

where

$$\overline{\omega}_{\mu mn} = \omega_{\mu mn} + (i\kappa^2/4)(\overline{\psi}_\mu \gamma_m \psi_n + \overline{\psi}_m \gamma_\mu \psi_n - \overline{\psi}_\mu \gamma_n \psi_m)$$

and

$$\omega_{\mu mn} = (1/2)e_m^\nu(\partial_\mu e_{n\nu} - \partial_\nu e_{n\mu}) + (1/2)(e_m^\rho e_n^\sigma \partial_\sigma e_{\rho p} e_\mu^p) - (m \leftrightarrow n)$$

($e_m^\nu$ is the verbein--the 'square root' of the metric tensor, $\omega_{\mu mn}$ is the spin connection--in this context representing the rotation of the gravitino--and $\sigma^{\mu\nu}$ is the spin generator, the generator of the spinor representation SU(2)XSU(2) of the Lorentz group.)

Note that the action $\int d^4x \, \varepsilon^{\mu\nu\rho\sigma} \, \overline{\psi}_\mu \gamma_5 \gamma_\nu \overline{D}_\rho \Psi_\sigma$ is just the massless Rarita-Schwinger action $\int d^4x \, \varepsilon^{\mu\nu\rho\sigma} \, \overline{\psi}_\mu \gamma_5 \gamma_\nu \partial_\rho \Psi_\sigma$ — the kinetic term for the gravitino, where the ordinary derivative has been replaced by the covariant derivative (2). The Einstein action corresponds to the action of the graviton, so that the action (1) is supersymmetric. The action S is also covariantized; i.e. locally supersymmetric; i.e. invariant under the action of order $\kappa^2$ due to the replacement $\partial_\rho \rightarrow \overline{D}_\rho$, which describes the rotation in spacetime of the graviton—the gauge field of local supersymmetry [J. Bailen, 1994]. Note that the derivative (2) differs from the ordinary gauge-covariant derivative. In ordinary gauge theory, the generators $T^A{}_{\alpha B}$ : A, B = 1,2,…,N; $\alpha$ = 1,…,$N^2$-1 represent orientations that constitute the fundamental representation of the gauge group. In local supersymmetry the generators $\sigma^{mn}$ represent orientations in spacetime.

The action (1), which derives from the gauging of global supersymmetry, generates a graviton vertex operator

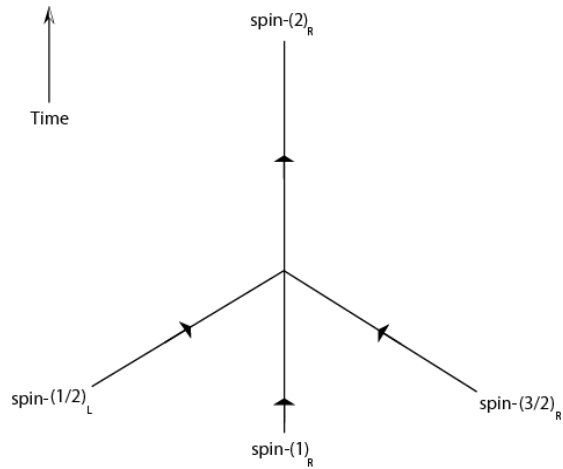

Figure 1

which can also occur in the alternative form

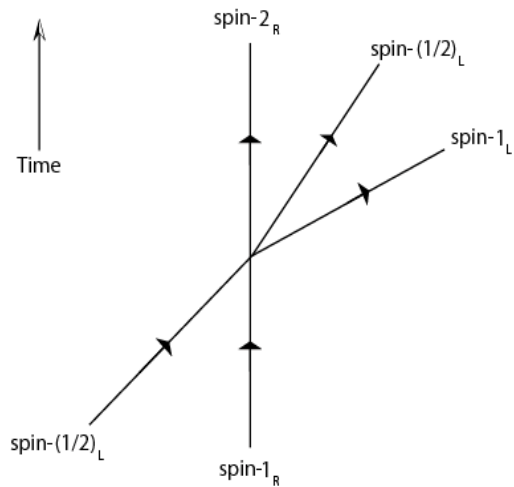

Figure 2

The Figure 2 alternative to the Figure 1 vertex operator is clearly generated by the replacement of the ingoing, right-handed gravitino with an outgoing fermion-boson pair--both fermion and boson of left-handed helicity. A remarkable aspect of this version of the graviton vertex operator is that if the ingoing fermion is a lepton (quark) and if the outgoing fermion is a quark (lepton), then the spin-2 field that is generated is always right-handed (left-handed) and of charge -2/3 (2/3).

The Figure 2 alternative to the traditional graviton vertex operator will serve as the tree-level prototype for the interactions that produce quark-lepton transitions. Three examples of such transitions are as follows:

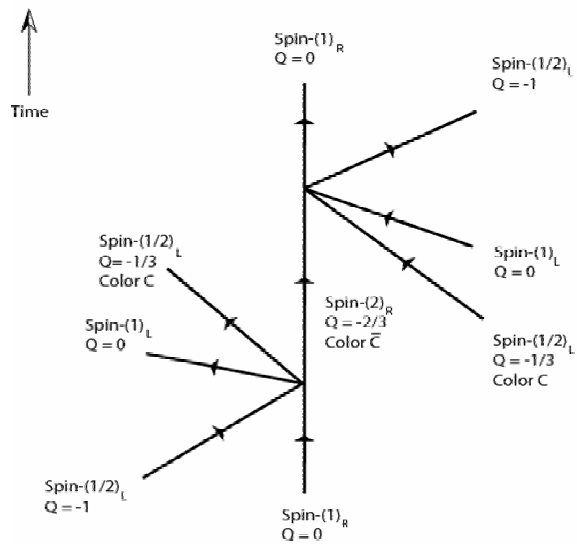

Figure 3

Here a right-handed boson absorbs a left-handed fermion of charge -1, which combination simultaneously radiates a left-handed fermion of charge -1/3 and of color C (which, due to its quantum numbers, is identified as a down quark), a left-handed boson of charge zero, and a spin-2 field of right-handed helicity, charge –2/3 and the anti-color of C, identified simply as '$\overline{C}$.' The spin-2 field absorbs a left-handed boson of charge zero, and a left-handed fermion of charge -1/3 and color C (a down quark), and immediately radiates a right-handed boson of charge zero and a left-handed fermion of charge –1 which, due to its quantum numbers, is identified as an LH electron. Note that charge, supersymmetry and color are conserved by the depicted vertices.

A second realization of the generic, Figure 2 interaction involves a right-handed boson that absorbs a left-handed fermion of charge zero (identified as the LH electron's neutrino):

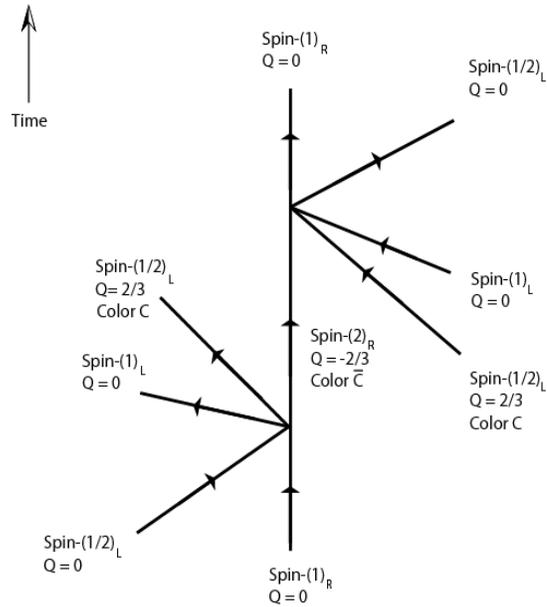

Figure 4

This combination simultaneously radiates a left-handed fermion of charge 2/3 and color C (identified as an up quark), a left-handed boson of charge zero and a spin-2 field of right-handed helicity, charge −2/3 and the anti-color $\overline{C}$ of C. Secondly, the spin-2 field simultaneously absorbs a left-handed boson of charge zero and a left-handed fermion of charge 2/3 and color C (an up quark), and radiates a right-handed boson of charge zero and a left-handed fermion of charge zero, which due to its quantum numbers, is identified as the LH electron's neutrino.

    A third realization of the generic Figure 2 interaction involves a left-handed boson that absorbs a right-handed quark of charge −1/3 and color C:

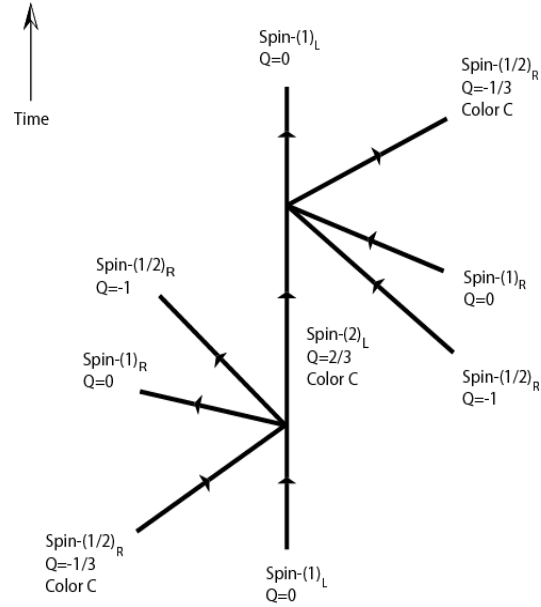

Figure 5

This combination simultaneously radiates a left-handed spin-2 field of charge 2/3 and color C, a right-handed boson of charge zero and a right-handed fermion of charge –1, which, due to its quantum numbers, is identified as an RH electron. The spin-2 field absorbs a right-handed boson of charge zero and a right-handed fermion of charge –1 which is identified as an RH electron; and radiates a left-handed boson of charge zero and a right handed fermion of charge –1/3 and color C, which is identified as a strange quark.

## II. Quarks as Gravitationally Excited Leptons

The above transitions are interesting because Figures 3, 4 and 5 indicate that every quark can be represented as a lepton that has absorbed a charge of 2/3. Specifically, assigning $I_3(e^-_L) = -1/2$ and $I_3(\nu^{e^-}) = +1/2$, the hypercharges ($Y = 2Q - 2I_3$) of the light quark generation and the strange quark are given by

$$Y(up) = 2[Q(\nu^{e^-}) + (2/3)] - 2I_3(\nu^{e^-}) = 2\{0 + 2/3\} - 2(1/2) = 1/3 \qquad (3A)$$

$$Y(down) = 2[Q(e^-_L) + (2/3)] - 2I_3(e^-_L) = 2(-1 + 2/3) - 2(-1/2) = 1/3 \qquad (3B)$$

$$Y(strange) = 2[Q(e^-_R) + (2/3)] - 2I_3(e^-_R) = 2(-1 + 2/3) - 2(0) = -2/3 \qquad (3C)$$

Thus, if the color that characterizes the quarks of expressions 3 is accounted for in terms of a new principle of equivalence that associates supergravity interactions with the introduction of color (as indicated in Figures 3, 4 and 5), then every quark can be interpreted as a gravitationally excited lepton.

By symmetry, the above construction is also applicable to the heavier generations of quarks.

$$Y(top)=2[Q(\nu^{\tau-})+(2/3)]-2I_3(\nu^{\tau-})=2\{0+2/3\}-2(1/2)=1/3 \qquad (4A)$$

$$Y(bottom)=2[Q(\tau^-_L)+(2/3)]-2I_3(\tau^-_L)=2(-1+2/3)-2(-1/2)=1/3 \qquad (4B)$$

and

$$Y(charmed)=2[Q(\nu^{\mu-})+(2/3)]-2I_3(\nu^{\mu-})=2\{0+2/3\}-2(1/2)=1/3 \qquad (5A)$$

$$Y(?)=2[Q(\mu^-_L)+(2/3)]-2I_3(\mu^-_L)=2(-1+2/3)-2(-1/2)=1/3. \qquad (5B)$$

Although expressions 4 and 5 represent a departure from traditional theory in the sense that top, bottom and charmed quarks are characterized by the quantum numbers $I_3$ and Y [c.f. D. Nordstrom, 1992], symmetry seems to argue for the validity of this construction.

Note that expression 5B indicates a quark that is not currently recognized. The strange quark is usually regarded as the generational partner of the charmed. But the strange quark cannot be inserted here (the hypercharge of strange is -2/3). Thus, the expression

$$Y(?) = 2[Q(\mu^-_L) + (2/3)] - 2I_3(\mu^-_L) = 2(-1 + 2/3) - 2(-1/2) = 1/3.$$

predicts a new quark—a quark characterized by the same quantum numbers as the strange quark, except that the new quark is left-handed. To avoid violation of six quark—six lepton symmetry, the predicted quark is regarded as a left-handed (non-strange) version of the strange quark [J. Towe, 1997].

Note that the postulated equivalence of supergravity and color is not surprising because the postulated increment of color produces an increment of 4-curvature

$$F^\alpha = [D,D]A^\alpha = (\partial_\mu A_\nu^\alpha - \partial_\nu A_\mu^\alpha)dx^\mu \wedge dx^\nu + iC^\alpha{}_{\beta\gamma}A_\mu^\beta A_\nu^\gamma dx^\mu \wedge dx^\nu$$

($D_\mu$: $\mu = 0,1,2,3$ is the color SU(3)-covariant derivative, the $A^\alpha = A^\alpha_\mu dx^\mu$ : $\alpha = 1,\ldots,n^2-1$ constitute the color connection and the $C^\alpha{}_{\beta\gamma}$ are the Cartan structural constants), which, incidentally, can occur in the absence of mass (an important consideration since supersymmetry theories can be massless); i.e. an increment of 4-curvature that is associated with the Rarita-Schwinger action—which corresponds to second order in $\kappa$.

### III. An Irreducible Representation of SUSY SU(5) In Terms of Leptons

Given the hypothesis that every quark is a lepton that has absorbed a graviton of charge 2/3), one can exactly constitute a minimal irreducible representation $5 \oplus 10$ of SUSY SU(5) in terms of three generations of leptons and scaler superpartners. The strictly leptonic realization of the anti-symmetric $10 = [5, 2]$ of SUSY SU(5) can be given by

$$10^{\text{LEP}} = \begin{pmatrix} 0 & e^-_L & v^{e-}_L & \tau^-_L & v^{\tau-}_L \\ -e^-_L & 0 & e^-_R & \tau^-_R & S_{e-} \\ -v^{e-}_L & -e^-_R & 0 & S_{\tau-} & S_{v(\tau)} \\ -\tau^-_L & -\tau^-_R & -S_{\tau-} & 0 & S_{v(e)} \\ -v^{\tau-}_L & -S_{e-} & -S_{v(\tau)} & -S_{v(e)} & 0 \end{pmatrix}. \qquad (6A)$$

The strictly leptonic realization of the symmetric $5 = [5, 1]$, complementing $10^{\text{LEP}}$, can be given by

$$5^{\text{LEP}} = \begin{pmatrix} \mu^-_L \\ v^{\mu-}_L \\ \mu^-_R \\ S_{\mu-} \\ S_{v(\mu)} \end{pmatrix}. \qquad (6B)$$

The realization of the anti-symmetric $10 = [5, 2]$ in terms of quarks can be given by

$$10^{\text{QRK}} = \begin{pmatrix} 0 & d & u & b & t \\ -d & 0 & s, S_B & s, S_A & S_d \\ -u & -s, S_B & 0 & S_b & S_t \\ -b & -s, S_A & -S_b & 0 & S_u \\ -t & -S_d & -S_t & -S_u & 0 \end{pmatrix}. \qquad (7A)$$

where the components (s, $S_A$) and (s, $S_B$) respectively represent the simultaneous production of a strange quark and a boson that has absorbed a scalar $S_A$; and the simultaneous production of a strange quark and a boson that has absorbed a scaler $S_B$. The realization in terms of quarks of the symmetric 5 of SU(5) can be given by

$$5^{\text{QRK}} = \begin{pmatrix} ? \\ c \\ s \\ S_s \\ S_c \end{pmatrix}, \qquad (7B)$$

where the particle designated '?' represents the new, predicted quark.

Note that each component $10^{\text{LEP}}_{ij}$: $i,j = 1,2,3,4,5$ is transformed into its counterpart $10^{\text{QRK}}_{ij}$ by the tree level interactions of Figures 3, 4 and 5, and that each component $5^{\text{LEP}}_i$: $i = 1,2,3,4,5$ is transformed into its counterpart $5^{\text{QRK}}_i$ by these interactions, so that the proposed irreducible representation $5 \oplus 10$ of SUSY SU(5) is invariant under the proposed supergravity interactions.

### IV. Impact Upon the Hierarchy Problem

The transitions that are described by Equations 5B and are depicted by Figure 5 are objectionable in the context of traditional theory, because the interaction involving an RH electron and an LH boson produces a strange quark, whether the right-handed particles are $e^-_R$, $\mu^-_R$ or $\tau^-_R$. To solve this problem, one first considers the chiral degrees of freedom that are provided by the lepton and quark realizations of the irreducible representation $5 \oplus 10$ of the proposed SUSY SU(5). The lepton realization of $5 \oplus 10$ precisely accommodates the chiral modes that are represented by the three generations of left and right handed electrons and neutrinos and their scaler superpartners. But the chiral modes that correspond to the quark realization of this representation do not exhaust the available degrees of freedom unless they include the modes that associate with the scaler particles (identified in equation 7A as) 'scaler A' and 'scaler B.' In this context, the interaction that produces a strange quark from an $e^-_R$ can be understood in terms of a hypothesis that the LH boson of Figure 5 has absorbed a scaler B, of mass-scale $10^{-1}$ GeV; and the interaction that produces the strange from a $\tau^-_R$ can be understood in terms of a hypothesis that the LH boson of Figure 5 has absorbed a scaler A, of mass-scale $10^2$ GeV.

This hypothesis also provides an explanation of the two mass scales $M_X$ and $M_Y$ that characterize SU(5) GUT theories. In the proposed theory, these mass scales are due to the difference between the lepton and quark realizations of $5 \oplus 10$. Because the difference $\Delta M_{X-Y} = M_X - M_Y \approx 10^{13}$ between the mass scales $M_X \approx 10^{16}$ GeV and $M_Y \approx 10^3$ GeV is so large, $\Delta M_{X-Y}$ appears unrelated to $\Delta M_{A-B} = M_A - M_B$. But the proposed theory argues that that $\Delta M \equiv \Delta M_{X-Y} = \Delta M_{B-A}$ is a running hierarchy that depends upon the kinetic energy level per particle. Specifically, it is postulated that the general hierarchy

$$\log(KE) + \log(\Delta M) = \log(M_X) \qquad (i)$$

is a constant that relates the ultra-high energy scale to the macro-scale. It is observed that if the kinetic energy per particle is 1 TeV ($10^3$ GeV)—the scale at which electroweak symmetry is broken, then $\Delta M = M_X - M_Y$ equals $10^{13}$ GeV, as when fine-tuned in conventional theory, so that the sum $\log(KE) + \log(\Delta M) = 16$ ($M_X = 10^{16}$ GeV). Secondly, it is hypothetically assumed that $\Delta M = 10^3$ GeV, as in the supergravity interactions described by Figure 5. In this context, one can maintain a constant mass-scale $M_X$ by describing the kinetic energy level of the proposed supergravity interactions as $10^{13}$ GeV, which appears reasonable.

The above discussion is clarified by considering the interactions that are beyond tree-level; i.e. the interactions that involve high-energy radiative corrections to the probability amplitude for an interaction that is mediated by the bosons under consideration. For interactions beyond tree-level, one would presumably obtain a total energy

$$\log(KE) + \log(\Delta M) + \log(RC) = \log(M_X), \qquad (ii)$$

where RC designates energy imparted by radiative corrections. It had already been postulated that the hierarchy represented by (i) is a constant. The term log(RC) clearly spoils the postulated constant status of (i), which is only cured (the term log(RC) eliminated) by the involvement of supersymmetry. In a context of supersymmetry then, (ii) can essentially be reduced to (i). But (i) indicates that, in principle, a high kinetic energy can diminish or 'spoil' traditional hierarchy ΔM, just as can radiative corrections. Clearly, this 'spoiling of hierarchy' cannot be cured by supersymmetry. It can however be cured by replacing the traditional hierarchy ΔM by the more general hierarchy (i), which is postulated as constant.

## References

1. J. Bailin and A. Love, 1994, *Supersymmetric Gauge Field Theory and String Theory*, Bristol, Institute of Physics Publishing, Bristol and New York, 299.

2. *Physical Review D, Particles and Fields*, Eds. D. Nordstrom, L. Brown and- S. Brown (1992, AIP Press, New York), III.68.

3. J. Towe, 1997, *Flavor Changing Neutral Currents: Present and Future Studies*, Ed. D. Cline (World Scientific, Singapore) 228.

## Figures

1. The Graviton Vertex Operator

2. A Modified Vertex Operator

3. Down Quark from LH Electron

4. Up Quark from Electron's Neutrino

5. Strange Quark from RH Electron